\documentclass[11pt, a4paper, twocolumn]{article}  
\usepackage{graphicx}
\usepackage{amsmath}
\usepackage{cite}
\usepackage{url}
\usepackage{enumitem}

\newcommand{\Leff}{L^{\prime}}

\title{The sound of the udu}
\author{Leonardo Pereira Vieira$^1$, Carlos Eduardo Aguiar$^2$ 
\\[4pt]
$^1$ {\small Centro Federal de Educa\c c\~ao Tecnol\'ogica Celso Sukow da Fonseca, RJ, Brasil} \\ 
$^2$ {\small Instituto de F\'\i sica, Universidade Federal do Rio de Janeiro, RJ, Brasil}}
\date{}

\begin{document}

\maketitle
\begin{abstract}
The udu is a traditional Nigerian percussion instrument that can be thought of as a Helmholtz resonator with two apertures. We use a simple two-particle mechanical analogue to study the behaviour of such a resonator and find that the model describes quite well the main features of udu acoustics.
\end{abstract}

{\it Keywords\/}: udu, Helmholtz resonator  

\section{Introduction}
\label{intro}

\begin{figure}[htb]
\centering
\includegraphics[width=0.7\linewidth]{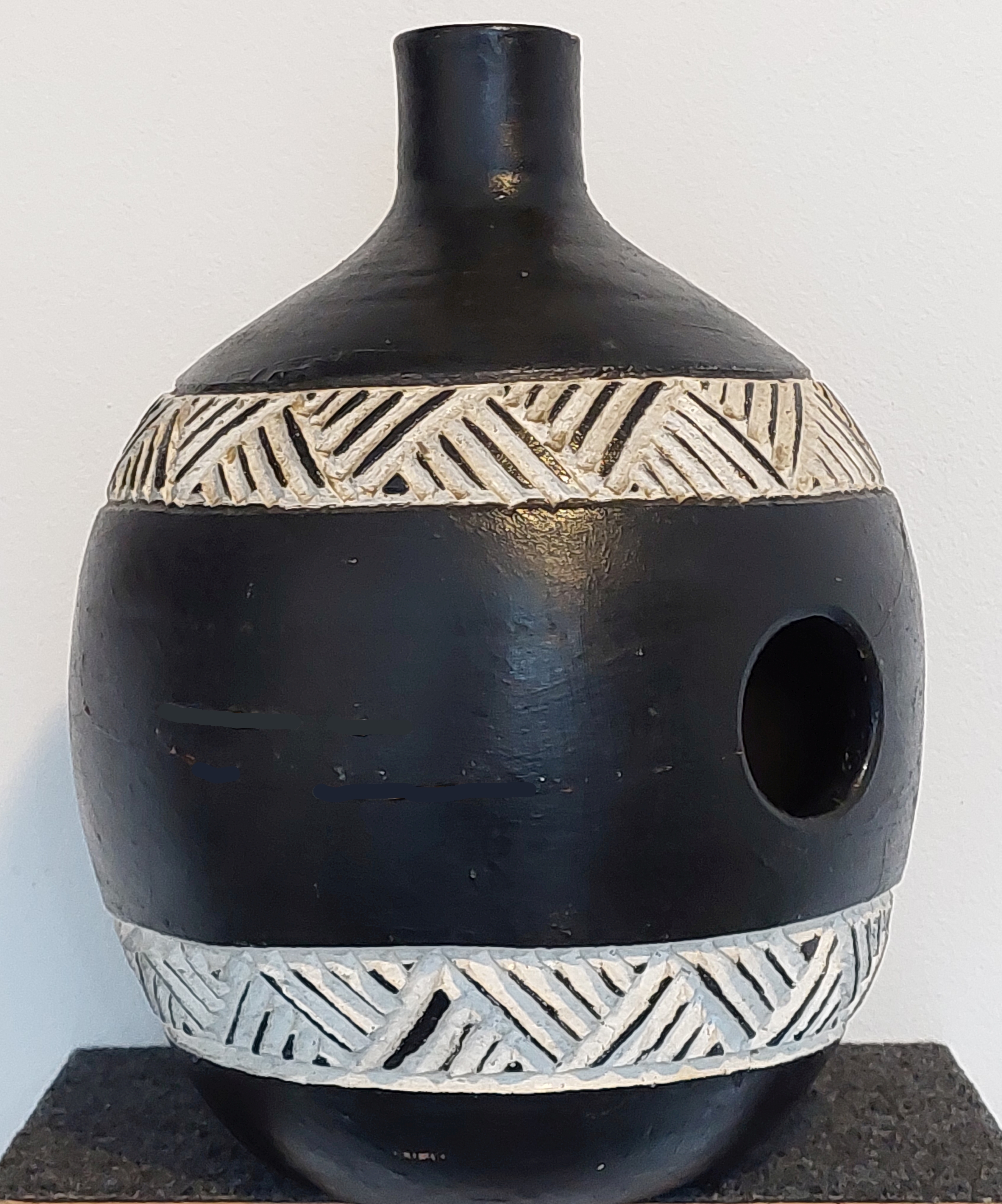}
\caption{The udu used in this work.}
\label{udu}
\end{figure}

The udu is a hand percussion instrument, a `clay drum' originating from the Igbo people in Nigeria, where it is traditionally played by women. 
Udu means vessel in Igbo language and the instrument has the shape of a water pot with an extra opening on the side, as shown in figure \ref{udu}. 
The udu produces a peculiar sound (sometimes described as `watery') when the body or neck openings are hit with the palm of the hand. 

As we will see, the udu is basically a Helmholtz resonator with two openings, and its characteristic sound can be understood in terms of a simple mass-spring-mass mechanical analogue in which the two-body reduced mass plays an interesting role. This makes the udu an attractive topic to introductory courses on vibrations and waves, allowing for theoretical and experimental explorations.
In section~\ref{resonators} of this paper we briefly describe the standard Helmholtz resonator and its two-aperture version, together with their mechanical analogues. 
Measurements of the main features of the udu sound are presented in section~\ref{sound} and shown to be quite well described by the mechanical model. In section~\ref{conclusion} we make some concluding remarks.

\section{Helmholtz resonators}
\label{resonators}

\subsection{Resonator with one aperture}
\begin{figure}[htb]
\centering
\includegraphics[width=0.65\linewidth]{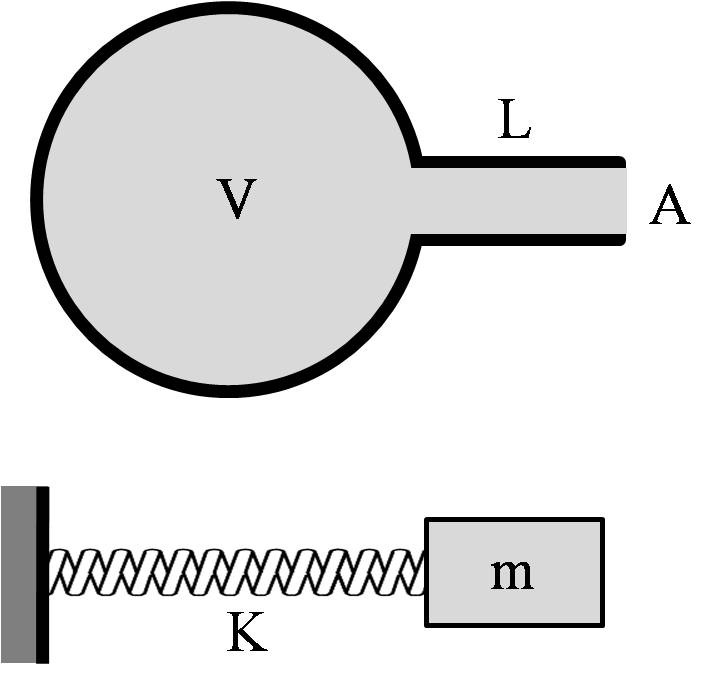}
\caption{A Helmholtz resonator (top) and its mechanical analogue (bottom).}
\label{onehole}
\end{figure}

The usual Helmholtz resonator~\cite{Greenslade, Dosch1, Kinsler} is a vessel containing air, connected to the exterior atmosphere by a slender neck as depicted in figure~\ref{onehole}~(top). The main cavity of the resonator has volume $V$ and can be of any shape. We will consider the neck to be a cylinder of length $L$ and cross-sectional area $A$.

Acoustic vibrations cause the air to move into and out of the cavity, changing the internal pressure $p$. If a small volume $\eta$ of air leaves the cavity, the adiabatic change in pressure relative to the equilibrium value is 
\begin{equation}
\Delta p = \frac{\partial p}{\partial V} \eta 
               = - \frac{\gamma p}{V} \eta \,, 
\end{equation}
where $\gamma$ is the conventional ratio of specific heats. To this pressure difference one can associate a potential energy $ U = - \int \Delta p \, d\eta $, or
\begin{equation}
 U = \frac{1}{2} \frac{\gamma p}{V} \eta^2 .  
\end{equation}
A kinetic energy $T$ can also be defined for this system. Assuming that the column of air moving in the neck has mass $\rho A L$ ($\rho$ is the air density) and velocity $\dot{\eta}/A$, we have 
\begin{equation}
 T = \frac{1}{2} \frac{\rho L}{A} \dot{\eta}^2 .
 \label{kinetic}  
\end{equation}
A better estimate of the kinetic energy is obtained substituting the neck length $L$ in~\eqref{kinetic} with the effective length $\Leff$ of the moving column of air. If the opening has a circular cross-section of radius $R$, this is given as~\cite{Dosch1, Kinsler}
\begin{equation}
 \Leff = L + 1.4 R  
 \label{unflanged}
\end{equation}   
for an unflanged neck termination and
\begin{equation}
 \Leff = L + 1.7 R  
 \label{flanged}
\end{equation}  
for a flanged termination. 

The total energy of the system is, accordingly,
\begin{equation}
E = T + U = \frac{1}{2} m \dot{\eta}^2 + \frac{1}{2} K \eta^2 \,,
\label{energy}
\end{equation}
where we have introduced the acoustic mass $m$ (also called `inertance'~\cite{Kinsler}) and acoustic stiffness $K$, defined as
\begin{eqnarray}
  m = \rho \Leff / A \,, \label{m} \\
  K = \gamma p/V \,. \label{K}
\end{eqnarray}
From the energy equation \eqref{energy} we see that the Helmholtz resonator is a harmonic oscillator of frequency 
\begin{equation}
 f = \frac{1}{2\pi} \sqrt{\frac{K}{m}}  \,.
\end{equation}
Using equations \eqref{m} and \eqref{K} this can be written 
\begin{equation}
 f = \frac{1}{2\pi} \sqrt{\frac{\gamma p}{\rho}\frac{A}{V \Leff}}
     = \frac{c}{2\pi} \sqrt{\frac{A}{V \Leff}}  \,,
\end{equation}
where $c = \sqrt{\gamma p/\rho}$ is the speed of sound in air.

One may also think of the Helmholtz resonator in terms of the spring-mass analogue introduced by Rayleigh~\cite{Rayleigh} and shown in figure~\ref{onehole} (bottom). The `mass' $m$ and `stiffness' $K$ of the mechanical system are the same as in the acoustic case, given by equations~\eqref{m} and~\eqref{K}, so that the analogue has the same oscillation frequency as the resonator.  
We will see next that models like this make it easier to understand more complicated Helmholtz resonators such as the udu.

\subsection{Resonator with two apertures}
\begin{figure}[htb]
\centering
\includegraphics[width=0.8\linewidth]{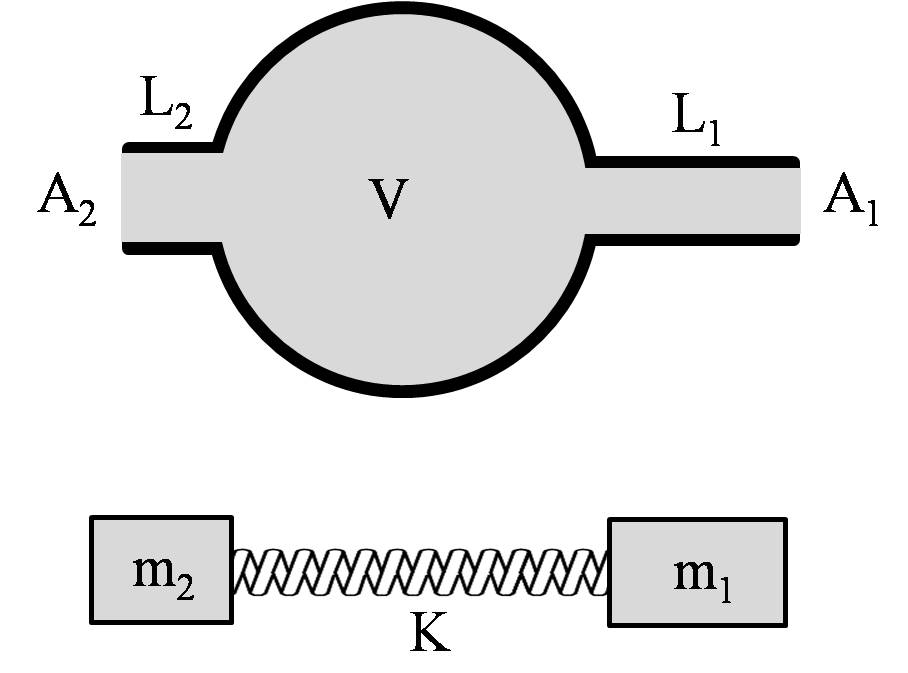}
\caption{A Helmholtz resonator with two openings (top) and its mechanical analogue (bottom).}
\label{twohole}
\end{figure}

The udu can be considered a Helmholtz resonator with two apertures, represented schematically in figure~\ref{twohole} (top). The main cavity of this resonator has volume $V$, as before. The necks, assumed to be well separated, have lengths $L_1$, $L_2$ and circular cross-sectional areas $A_1$, $A_2$.

A mechanical analogue of the double-aperture Helmholtz resonator is the mass-spring-mass system shown at the bottom of figure~\ref{twohole}, a straightforward extension of the model for the one-aperture resonator. Similarly to the previous case, the stiffness of the analogue spring is determined from the cavity volume via equation \eqref{K}, and masses $m_1$, $m_2$ are given by equation \eqref{m} with the corresponding values of the neck area and effective length substituted.
A standard analysis of the relative motion of the two masses~\cite{Kibble} then shows that the oscillation frequency, and consequently the frequency of the double-aperture Helmholtz resonator, is 
\begin{equation}
   f = \frac{1}{2\pi} \sqrt{\frac{K}{\mu}}  
   \label{freq}
\end{equation}
where $\mu$ is the reduced mass, given by
\begin{equation}
   \frac{1}{\mu} = \frac{1}{m_1} + \frac{1}{m_2}   \,.
   \label{mu}
\end{equation}

From equations \eqref{freq} and \eqref{mu}, one immediately derives an interesting consequence of the two-body nature of the mechanical analogue.  Because the reduced mass is the relevant inertial parameter, we find a relation already known to Helmholtz and Rayleigh~\cite{Rayleigh},
\begin{equation}
   f  = \sqrt{f_1^2 + f_2^2} 
   \label{pitagoras} 
\end{equation}
where
\begin{equation}
 f_i = \sqrt{K / m_i}  \,, \quad i=1,2
 \label{fi}
\end{equation}
is the frequency of the system when mass $m_i$  oscillates at one end of the spring while the other mass is clamped. For the double-aperture resonator this means hole $i$ is open and the other hole is closed.
In the next section we compare all these results to the frequencies measured on different beats of a udu.

\section{The udu sound}
\label{sound}

When playing the udu, one may hit the openings with the palm of the hand in various ways, creating different sounds. Identifying the neck and body holes as apertures 1 and 2 of the previous section, we have possibilities like:
\begin{enumerate}[label=(\roman*)]
\item\label{i1} hit the body hole and keep it closed by the hand, leaving the neck hole open and producing sound of frequency $f_1$; 
\item\label{i2} hit the neck hole and keep it closed by the hand, leaving the body hole open and producing sound of frequency $f_2$; 
\item\label{i3}  hit one hole and quickly withdraw the hand, so that both holes are effectively open, producing sound of frequency $f$. 
\end{enumerate}

\begin{figure}[htb]
\centering
\includegraphics[width=0.95\linewidth]{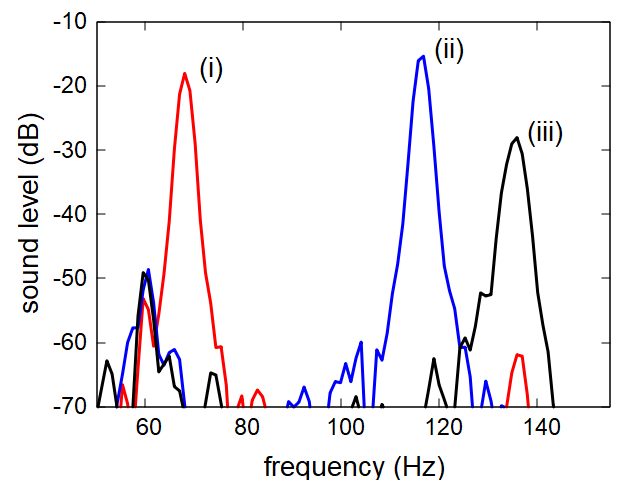}
\caption{Frequency spectra of the udu for different beats: \ref{i1} only neck hole open,  \ref{i2} only body hole open, \ref{i3} both holes open.}
\label{spectra}
\end{figure}

We have recorded the sound of beats~\ref{i1} to~\ref{i3} on the udu shown in figure~\ref{udu}. The waveforms were Fourier analysed with the Audacity audio software~\cite{Audacity} and the resulting frequency spectra are presented in figure~\ref{spectra}. Each of the three spectra displays a single prominent peak, at frequencies $f_1=68$~Hz for beat~\ref{i1}, $f_2=116$~Hz for beat~\ref{i2} and $f=136$~Hz for beat~\ref{i3}.

The two-body character of the udu dynamics can be tested by checking how well relation~\eqref{pitagoras}, derived from the reduced mass, is fulfilled. From the measured values of $f_1$ and $f_2$,  this relation predicts that $f$ should be 134.5~Hz, in good agreement ($\sim 1\%$) with the experimental value 136~Hz.

Furthermore, the peak frequencies of the three beats are quite well reproduced by the Helmholtz resonator formulas~\eqref{freq} and~\eqref{fi}. The volume of the udu cavity was measured by filling it with a known quantity of water and found to be $ V = 10.2$~litres. The neck length and internal radius are respectively $ L_1 = 6.3$~cm and $R_1 = 2.2$~cm, so that the effective length is $\Leff_1 = 9.4$~cm according to equation~\eqref{unflanged}. 
The body aperture has length $L_2 = 1.2$~cm (the wall thickness) and radius $R_2 = 3.0$~cm. Regarding the outer surface as a flanged termination to the aperture and using equation~\eqref{flanged}, the effective length is $\Leff_2 = 6.3$~cm.
With these dimensions and considering the speed of sound is $c = 340$~m/s, the calculated frequencies have the values given in table~\ref{comp}. The measured frequencies are also shown in the table in order to facilitate comparison. Again, we find good agreement (to a few percent) between the model and experimental results.
\begin{table}[htb]
\centering
  \begin{tabular}{ l c c c }
    \hline
                    & $f_1$ (Hz) & $f_2$ (Hz) & $f$ (Hz) \\ \hline
    Calculated  & 68            & 113          & 132       \\ 
    Measured   & 68            & 116          & 136       \\
    \hline
  \end{tabular}
\caption{Calculated and measured frequencies of the udu beats.}
\label{comp}
\end{table}

\begin{figure}[htb]
\centering
\includegraphics[width=0.95\linewidth]{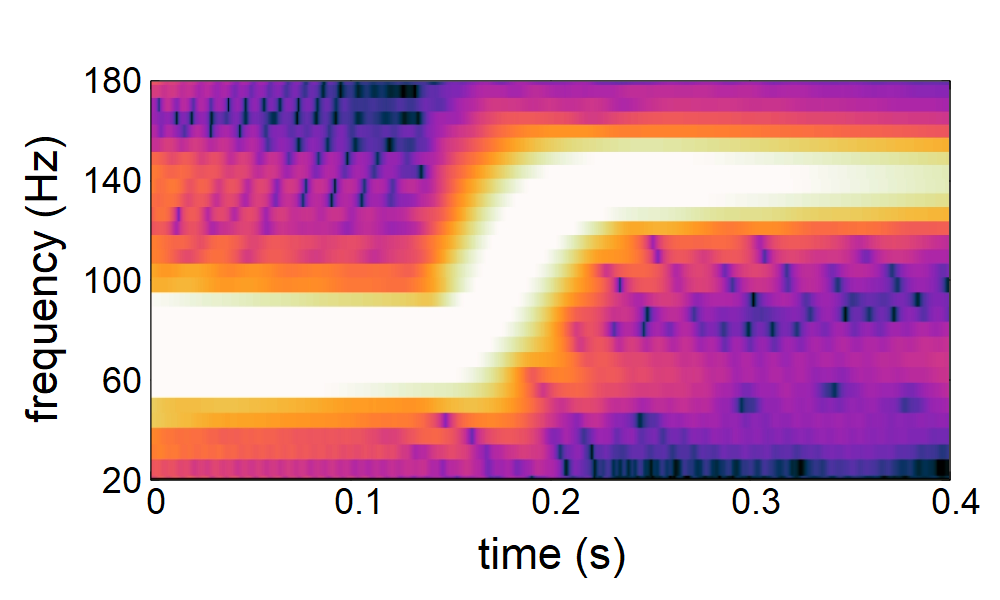}
\caption{Spectrogram of the sound produced by hitting the udu body hole with the hand, keeping it closed for some time and then releasing it. The bright band indicates the peak frequency as function of time.}
\label{spectrogram}
\end{figure}

We can also analyse a very distinctive udu sound, produced by hitting the body hole with the palm of the hand, keeping it closed for some (small) time and then taking away the hand. A spectrogram of this beat is shown in figure~\ref{spectrogram}, exhibiting a transition from low to high frequency (from $f_1$ to $f$) that makes it sound like a droplet falling on water.
This change to a higher frequency is essentially a reduced mass effect.  Opening the body hole is equivalent to releasing a clamped mass, $m_2$ in the case, shifting from a one-body to a two-body oscillator. 
The reduced mass of two bodies is necessarily smaller than any of the individual masses, so that the transition in this beat is always from a lower to a higher pitch, as observed.

\section{Concluding remarks}
\label{conclusion}

Musical instruments are popular subjects in courses on vibrations and waves. We have seen in this paper that a traditional Nigerian drum, the udu, should join the list of instruments that are fruitfully discussed in such courses. The main characteristics of the sound of a udu are easily understood in terms of a simple mechanical model---two masses connected to a spring---in which the reduced mass plays an interesting and perhaps surprising role. 
Quantitatively, the model predicts the frequencies of different udu beats to within a few percent of the measured values.

We should note that there are other musical instruments, like the ocarina, that operate as Helmholtz oscillators with multiple apertures. An extension of the present study to such instruments might be worth pursuing, not only because their sound can be described with relatively simple models, but also for the interest they usually spark on students.


\begin{thebibliography}{99}

\bibitem{Greenslade} T.~B.~Greenslade, Experiments with Helmholtz resonators, {\it The Physics Teacher} {\bf 34}(4), 228--230 (1996).

\bibitem{Dosch1} H.~G. Dosch and M.~Hauck, The Helmholtz resonator revisited, {\it European Journal of Physics} {\bf 39}(5), 055801 (2018).

\bibitem{Kinsler} L.~E.~Kinsler, A.~R.~Frey, A.~B.~Coppens and J.~B.~Sanders, {\it Fundamentals of Acoustics}, 4th~ed. (Wiley, 2020) sec.~10.8--9. 

\bibitem{Rayleigh} J.~W.~Strutt (Lord Rayleigh),  {\it The Theory of Sound}, 2nd~ed., 1894 (Dover, 1969) ch.~XVI. 

\bibitem{Kibble} T.~W.~B.~Kibble and F.~H.~Berkshire, {\it Classical Mechanics}, 5th~ed. (Imperial College Press, 2004) ch.~7.

\bibitem{Audacity} Audacity Audio Software, 
\url{<https://www.audacityteam.org>}.

\end{thebibliography}
\end{document}